\documentclass[
superscriptaddress,
 amsmath,amssymb,
 aps,
]{revtex4-2}

\usepackage{amsmath}
\usepackage{dsfont}

\DeclareMathOperator*{\argmin}{arg\,min}

\usepackage{caption}
\usepackage{subcaption}

\usepackage{graphicx}

\usepackage{dcolumn}
\usepackage{bm}
\usepackage{hyperref} 
\usepackage{physics}

\usepackage{float}
\usepackage{xcolor}
\usepackage{comment}

\usepackage{mhchem}

\usepackage{url}

\usepackage{algpseudocode}

\usepackage{algorithm}
\usepackage{algorithmicx}

\algrenewcommand\algorithmicrequire{\textbf{Input:}}
\algrenewcommand\algorithmicensure{\textbf{Output:}}

\begin{document}

\preprint{APS/123-QED}

\title{Formulating Multistage Cutting Stock Problems as QUBO}

\author{Marcel Seelbach Benkner}
\affiliation{Department of Electrical Engineering and Computer Science, 
University of Siegen, Hölderlinstraße 3, 57076 Siegen, Germany
}
\affiliation{eleQtron GmbH, Heeserstraße 5, 57072 Siegen, Germany}

\author{Chiara Capecci}\affiliation{Pitaevskii BEC Center and Department of Physics, University of Trento, Via Sommarive 14,
38123 Trento, Italy}
\affiliation{INFN-TIFPA, Trento Institute for Fundamental Physics and Applications, Trento, Italy}

\author{Sebastian Nagies}\affiliation{Pitaevskii BEC Center and Department of Physics, University of Trento, Via Sommarive 14,
38123 Trento, Italy}
\affiliation{INFN-TIFPA, Trento Institute for Fundamental Physics and Applications, Trento, Italy}

\author{Javed Akram}\affiliation{eleQtron GmbH, Heeserstraße 5, 57072 Siegen, Germany}
\author{Sebastian Rubbert}\affiliation{eleQtron GmbH, Heeserstraße 5, 57072 Siegen, Germany}

\author{Dimitrios Bantounas}
\affiliation{eleQtron GmbH, Heeserstraße 5, 57072 Siegen, Germany}

\author{Philipp Hauke}\affiliation{Pitaevskii BEC Center and Department of Physics, University of Trento, Via Sommarive 14, 38123 Trento, Italy}
\affiliation{INFN-TIFPA, Trento Institute for Fundamental Physics and Applications, Trento, Italy}

\author{Michael Johanning}
\affiliation{eleQtron GmbH, Heeserstraße 5, 57072 Siegen, Germany}
\author{Michael Moeller}
\affiliation{Department of Electrical Engineering and Computer Science, 
University of Siegen, Hölderlinstraße 3, 57076 Siegen, Germany
}

\date{\today}

\begin{abstract} 
Cutting stock problems are of large relevance to a variety of industry branches. Here, we present a linear programming formulation for a restricted version of the  multistage 2D cutting stock problem with Guillotine cuts and test it on published benchmark instances. 
After this, we derive a non-exact reformulation in quadratic unconstrained binary optimization (QUBO) form via unbalanced penalization, which we show to have possible benefits in modeling the problem with usable leftovers. 
This work mainly considers the restricted formulation, in which the dimensions of each cut are determined by those of a single required piece. We further discuss how the approach can be extended to the unrestricted problem, which allows more general cutting patterns involving cuts with dimensions of multiple required pieces and naturally gives rise to quadratic terms in the inequalities of the problem formulation.
We showcase solutions via a Simulated Annealing solver for the restricted problem class, where we mainly study the performance of the iterative augmented Lagrangian method. 
We also discuss possible applications of Quantum Annealing in this setting, as a strong motivation to research QUBO formulations.

 \end{abstract}
\maketitle
\section{Introduction}
\begin{figure}
    \centering
\includegraphics[width=0.5\linewidth]{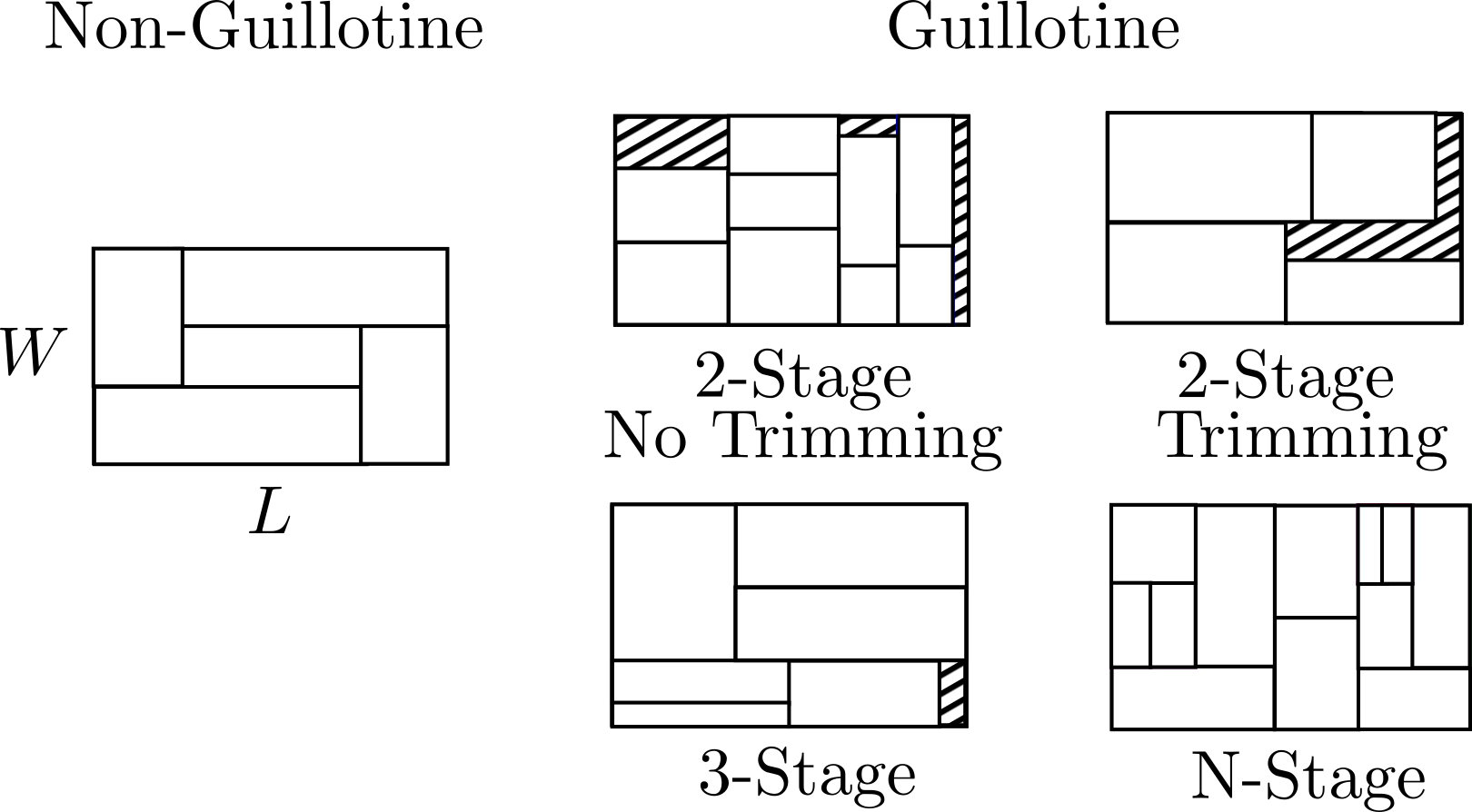}
    \caption{Different types of 2D cutting stock problems. Own visualization based on figure from \cite{haessler1991cutting}. Left: a plate of width $W$ and length $L$ is cut into several (here: 5 pieces), where any cuts are allowed. 
    Right: Real-world machines often work with so-called Guillotine cuts, where the plate is cut all the way from one to the other side. These can occur in various stages, where the pieces resulting from a previous stage are further cut into smaller pieces, again using Guillotine cuts.}
    \label{fig:GuillotineCuts}
\end{figure}

2D cutting stock problems deal with the task of most efficiently cutting rectangular pieces of certain sizes out of one or multiple larger plates. The problem has been researched at least since 1960 \cite{kantorovich1960mathematical} and is crucial in industry branches utilizing plywood, paper, metal plates, and glass sheets \cite{turbian2024two}. A common, technological restriction in cutting these materials is that it has to be done via Guillotine cuts. This means that one has to cut the material all the way through as it is depicted in Fig.~\ref{fig:GuillotineCuts}. The resulting pieces can be cut again, where one typically alternates between horizontal and vertical cuts. The number of times the direction is changed is named cutting stage. While traditionally such type of problems can be solved via mixed-integer linear programming, branch and bound methods, dynamic programming, or some heuristics \cite{becker2024comparative}, motivated by the growing interest in optimization via quantum annealing we investigate in this work the formulation as a quadratic unconstrained binary optimization problem (QUBO). Quantum annealing is a metaheuristic optimization method utilizing quantum mechanical systems \cite{Kadowaki1998,farhi2000quantum,hauke2020,mcgeoch2022adiabatic}. The rough idea is that via quantum mechanical tunneling some energy landscapes can be explored more efficiently than for example with classical simulated annealing \cite{das2005quantum}. Currently available quantum annealing hardware is mostly restricted to solving QUBO problems of small scale or with severe restrictions in the coupling matrix. Nevertheless, applications for optimization in industry are already investigated \cite{yarkoni2022quantum,DeAndoin2023,Slongo2023,Ghamari2022,meli2025quantum, nagies2026practical}.

The main goal of this work is to formulate the restricted multistage 2D cutting stock problem in a form suitable for QUBO-based optimization. Starting from a 0-1 integer linear programming formulation, we derive a QUBO formulation that handles inequality constraints without introducing slack variables. We investigate different strategies for enforcing these constraints, with particular focus on an iterative augmented Lagrangian method, and explore how the QUBO formulation can also be used to favor usable leftovers. We then benchmark the proposed approach, analyze its scaling, and investigate the quantum annealing dynamics of representative problem instances.

This paper is organized as follows. In the rest of this Introduction, we present quantum annealing and the MAGIC architecture as a reference trapped-ion platform on which the proposed approach could, in principle, be implemented experimentally. Section~\ref{sec:related_work} discusses related work on cutting stock problems, while Sec.\ref{sec:method} presents the linear and QUBO formulations and the methods used to solve them. Numerical results are presented in Sec.\ref{sec:exp}, followed by the scaling analysis in Sec.\ref{sec:scaling} and an analysis of representative QUBO instances through quantum annealing simulations in Sec.\ref{sec:Simulating}. Finally, Sec.\ref{sec:conclusion} presents our conclusions.

\subsubsection{Quantum Annealing}

Quantum annealing \cite{Kadowaki1998,farhi2000quantum,hauke2020,mcgeoch2022adiabatic} is a quantum computing protocol for solving optimization tasks. More specifically, an optimization problem is mapped onto a quantum-mechanical Hamiltonian that can be implemented in physical hardware. The optimization problem then becomes a search for the ground state of the Hamiltonian. The actual annealing process solves that by starting from a simple Hamiltonian with a known ground state, in which the system is prepared. Then, the Hamiltonian is continuously and slowly changed into the Hamiltonian with the ground state of interest. The underlying principle of this process is the adiabatic theorem \cite{Born1928}, which states that---in the absence of exact level crossings---a system in an instantaneous eigenstate of the Hamiltonian stays in that instantaneous eigenstate, as the Hamiltonian is continuously (and sufficiently slowly) transformed. 
For the annealing protocol to reach the final target ground state, it is therefore necessary to transform the Hamiltonian such that the ground state and the excited states do not encounter level crossings. Conversely, the minimum energy difference between the ground state and the excited states is one of the most important predictors of the required time for the annealing protocol \cite{farhi2000quantum}.

In combinatorial optimization problems, the final problem Hamiltonian is diagonal in the computational basis (in contrast to other applications, such as from quantum chemistry).  
The computational basis states then correspond to a solution candidate of the problem expressed in terms of decision variables. Examples for this could be the position of a cut, the assignment of items to a knapsack, or the choice of a certain shipping route \cite{nemhauser,kellerer2004multidimensional,papageorgiou2014mirplib}. 
To find the best solution, one therefore typically desires a final state that is a product state, which encodes the optimal solution~\footnote{A final superposition states, instead, deteriorate the solution quality, since the optimal solution in the final state has then a reduced probability amplitude~\cite{hauke2015probing,Santra2025,Capecci2025}. They can, however, be desired if one is interested in sampling from a degenerate manifold instead of in a single optimal solution~\cite{Slongo2023}.}. 
While initial and final state are therefore purely classical,   
in order to avoid the aforementioned level crossings, during the transformation of the Hamiltonian, terms that are non-diagonal in the computational basis are required. The standard choice for a full annealing protocol is to start from a Hamiltonian that is a sum over $\sigma_x$ terms on all qubits, and to linearly transform it into the Hamiltonian of interest. While this is the simplest choice, it is usually by far not the most effective. There has been tremendous interest in choosing annealing schedules and different paths through the Hamiltonian space \cite{guvery2019, takahashi2017, Sels2017, prielinger2021, passarelli2020, capecci2026quantum}.

Combinatorial problems often have a multitude of equivalent mathematical formulations. In classical discrete optimization, most problems are formulated as integer linear programs (ILP) or mixed integer linear programs (MILP), thereby leveraging highly performant standard solvers, such as Gurobi, CPLEX, or SCIP \cite{schrijver1998theory}. In quantum computing, including quantum annealing, the most common formulation of optimization problems are quadratic unconstrained binary optimization problems (QUBO)s \cite{glover2019, Lucas2014}. The absence of explicit constraints and the limit to quadratic and linear terms is naturally compatible with most real-world quantum hardware: Explicit constraints are not straightforward to implement using unitary transformations~\cite{Bottarelli2025}, since a constraint is naturally more similar to a projection. Limiting the model to second order terms is especially compatible with the hardware, since hardly any hardware has multi-qubit interactions that are natively implemented rather than parallelized into two-qubit interaction \footnote{Higher-order, or polynomial unconstrained binary optimization (PUBO), formulations are nevertheless worth considering, as the constant overhead of the decomposition into two-qubit interactions can be more than compensated by the ancillary qubits
saved and by the larger minimum energy gaps of the direct encoding~\cite{nagies2025boosting}.}.

\subsubsection{Quantum Optimization on ion traps with the MAGIC architecture}
\label{sec:PhysIntro}
Various quantum architectures lend themselves to implementing quantum annealing protocols \cite{King2023,deOliveira2026,Lu2025}. We review here a trapped ion quantum computing architecture \cite{lekitsch2017} based on magnetic gradient induced coupling (MAGIC) \cite{mintert2001ion, wunderlich2002conditional}, which has recently been proposed for quantum annealing applications~\cite{nagies2026practical}. In this architecture, trapped ions are subjected to a static magnetic field gradient along the direction of the ion chain. The qubits are encoded in low energy states of the ion, for example the $^2S_{1/2}, \,\, F=0$ and $F=1, \, m=\pm 1$ states in \ce{^{171}_{}Yb+}. Due to the non-zero magnetic quantum number, the qubit couples to the magnetic field. 
In the presence of a magnetic field gradient, the spatial variation of the magnetic field results in a position-dependent qubit resonance frequency. This position dependence has two major effects:

1. Each qubit in an ion chain has a unique addressing frequency, enabling individual control of the qubits with low crosstalk \cite{piltz2014trapped}, without the need of focusing a signal onto individual ions. 

2. The ion chain experiences slight deformations depending on the qubit states. This in turn causes the qubits to interact thanks to the non-linearity of the mutual Coulomb repulsion, hence the name magnetic gradient induced coupling. The resulting interaction is of all-to-all, always-on type, i.e., an interaction occurs between any two qubits at all times, without any external driving.  

While the position-dependent addressing frequency is central to MAGIC's concept for quantum control, it is possible to recode idle qubits using the $F=0$ and $F=1, m=0$ states, making the qubit insensitive to the magnetic field. This insensitivity permits to controllably remove a given ion from the interactions. It also decouples it to first order from magnetic field fluctuations, thereby making it a clock-qubit with significantly increased coherence time.

The effective Hamiltonian, ignoring phonon excitations, of an $N-$qubit register is:
\begin{align}
    H_{\text{\phantom{drive}}} &=  H_{\text{idle}} + H_{\text{drive}}, \\
    H_{\text{idle\phantom{e}}} &=  \frac{\hbar}{2} \left( \sum_{n=1}^N \omega_n \sigma_z^n - \sum_{i<j}^N J_{ij} \sigma_z^i \sigma_z^j \right), \\
    H_{\text{drive}} &=  \frac{\hbar \Omega(t)}{2} \sum_{n=1}^N \sigma_x^n,
\end{align}
where $\omega_n$ is the resonance frequency for qubit $n$, $\sigma_x$ and $\sigma_z$ are the respective Pauli matrices, $J_{i, j}$ is the coupling strength between qubits $i$ and $j$, and $\Omega(t)$ is the time dependent drive, technically implemented by a microwave source.  

$J_{i,j}$ only depends on typically fixed parameters, such as the magnitude of the magnetic gradient, the trapping potential of the ion chain, and the number of ions \cite{mintert2001ion, wunderlich2002conditional, johanning2009quantum, nagies2024role}. We achieve arbitrary effective coupling using gate synthesis \cite{piltz2016versatile}, whose underlying principle is as follows. The energy contributions from the two-qubit terms flip their sign if exactly one of the respective qubits undergoes a bit flip. Gate synthesis goes through multiple stages of bit flips followed by waiting times and reverses the bit flips in the end. The time average of the two-body energy terms yields the effective, synthesized interaction. 

The synthesized interaction enables us to implement any QUBO on a MAGIC device. Due to the nature of the synthetization, annealing will typically be carried out as a trotterized annealing approach, where the continuous transformation is approximated through a sequence of discrete steps \cite{nagies2026practical}.

\section{Related Work on the 2D Cutting Stock Problem}
\label{sec:related_work}
Because of the high industrial relevance, a lot of research to figure out how to best solve 2D cutting stock problems has been conducted. Reference \cite{becker2024comparative}, for example, contains a comparison between many different mathematical formulations.

A simplification to the 2D cutting stock problem that is important in this work and which has also been described in \cite{becker2024comparative,furini2013models} is that cuts always have to be of the length of one required piece. These problems are called restricted. Furthermore, cutting patterns in this work are even more simplified since intermediate pieces that arise during the cutting process will be a rest piece if they are not cut out with a length or width of a required piece.

In contrast to our method, multiple works consider only a limited number of cutting stages \cite{yanasse2006linear,furini2013models,andrade2016two,silva2010integer}. There exists also a QUBO formulation for two-stage cutting stock problems that is also motivated by using quantum annealing \cite{arai2021study}. In that work, required pieces are assigned to places at the base material. The number of performed cuts is the main objective for minimization.

Several works try to find optimal solutions to the related bin packing problem \cite{cellini2024qal, xu2025digitized}. The methodology of Ref.~\cite{cellini2024qal} is related to our work since it also investigates augmented Lagrangian methods. The formulation of the method that we use can be found in Ref.~\cite{sharma2025cutting}. The idea of augmented Lagrangian methods is to enforce constraints in an iterative way via a quadratic term, a method that is well known and widely used in the context of continuous optimization \cite{nocedal2006numerical, hestenes1969multiplier,powell1969method}.

Note that there is also work on how optimization problems formulated as MILPs can  be expressed as QUBOs (to arbitrary precision) or tackled in a hybrid way with QUBOs \cite{zhao2022hybrid}. In the simplest case, linear constraints are then enforced with slack variables and quadratic penalty terms.

Besides the problem to obtain pieces with the largest value on the plates we also consider the problem of having usable leftovers. In this scenario, it matters if the waste was cut into small pieces in the process or not. In fact, for our specific problem formulation we see generating usable leftovers as a main motivation for a reformulation as QUBO. 
The problem with usable leftovers was, for example, considered in \cite{do2022two} and in \cite{nascimento2025multi}. 
However, the benchmark instances that are utilized there typically have a high amount of required pieces with the same dimensions \cite{UsableLeftoversDataset}. Therefore, being able to solve problems with integers is a large benefit compared to using binary solvers. Instead, we will consider
some of the problem instances from 
\cite{fayard1998efficient}
and compare our result with their optimal solution.

\section{Method}
\label{sec:method}
\subsection{0-1 Integer Linear Programming Formulation}

Before we discuss useful QUBO formulations, we describe a 0-1 integer linear programming approach to solve the multistage restricted 2D cutting stock problem. 
To do this, we aim to build a tree structure with a binary solution vector. 
The nodes are the ready pieces and all the child nodes that are connected to the parent node describe the pieces that will be cut out from the same intermediate piece in later cutting stages. 
Cuts will always be applied alternately in the $x-$ and $y-$direction. 

We want to describe cutting plans with binary variables and set up problems to optimize the objective later.
The required pieces are referred to with indices $1,...,m, s_1,s_2 $, where $m$ is the total number of required pieces. They may also have the same dimensions and each has a value $v_i$. Binary variables $x_{k,l}$ describe if the piece $k$ will neighbor $l$ horizontally in some way, while binary variables $y_{k,l}$ describe if the piece $k$ will neighbor $l$ vertically in some way. Note that we do not track the exact position of ready pieces. Instead the cutting patterns are determined up to switching of parts that are generated with the same parent node.

We will refer to the indices $s_1, s_2$  as starting nodes. The binary variables involving $s_2$ are fixed so that only
$y_{s_1,s_2}=1$, while all the others are set to zero. From $s_2$, arbitrary pieces can be reached. These will determine the columns in the first cutting stage.

Every required piece should not appear more than once in the tree structure. 
For this reason, we have the inequalities 
\begin{equation}
\forall k \in \{1,...,N\} \quad \sum_j x_{j, k}+ y_{j, k} \leq 1, \label{eq:IneqDemand}
\end{equation}
with $N:=m+2$. If the same piece is demanded multiple times, one just lets multiple indices refer to required pieces with the same dimensions.
Since we are interested in the restricted problem, the cuts will have the size of one required piece and in each cutting stages new pieces will get ready. It is important that in the tree structure pieces are placed in a way that they are not bigger than the intermediate piece they are cut out from. To model this, we need the size constraints
\begin{equation}
 l_i \sum_t y_{t , i} +  \sum_{j} l_{j} x_{i, j} \leq 
 \sum_t l_t y_{t , i}  \label{eq:length}
\end{equation}
and 
\begin{equation}
 w_i \sum_t x_{t , i}  +  \sum_{j} w_{j} y_{i, j} \leq 
 \sum_t w_t x_{t , i}  \label{eq:width}
\end{equation}

Note that each piece $i$ can either be reached by a vertically or with a horizontally cut from a previous piece according to \eqref{eq:IneqDemand}. Therefore, for each piece one inequality should always have all occurring binary variables set to zero and prevent that two successive cutting stages all cut in the same direction. All variables where bigger pieces would be cut out from smaller pieces can be set to zero immediately. In order to not modify the size restriction inequalities for the starting pieces, we set $l_{s_2}=L$, $w_{s_1}=W$ and $l_{s_1}=0$.

To allow rotating pieces, we simply add a demanded piece $(w_{i+\text{start rot.}}, l_{i+\text{start rot.}})$ for each piece with measures $(l_i, w_i)$, where $\text{start rot.}$ is the index from which on all rotated pieces are labeled. This modeling has to be considered in the demand inequality constraint. For this, we change it to
\begin{equation}
    \forall k \in \{1,...,N\} \quad \sum_j \left( x_{j, k}+ y_{j, k} +x_{j, k+\text{start rot.}}+ y_{j, k+\text{start rot.}} \right)\leq1.
\end{equation}

Since everything is formulated as linear program one can just use a regular solver for the problem. 
In the case that there is enough space to place every part, the 0-1 integer linear programming formulation can have drawbacks. In practice, one wants to choose cutting patterns with large rest pieces that can be used in the future.
The QUBO model that we set up in the next section could possibly improve this.

\subsection{Reformulation as a QUBO}\label{sec:Ref}
We now reformulate the problem as a QUBO problem in a non exact way, which can also address the problem of having usable leftovers. 
First, we need to take care of inequalities \eqref{eq:IneqDemand}.
Fortunately, one does not need to use slack variables here, which is a common approach but which can incur significant additional overhead~\cite{Bottarelli2025}. Instead, one can just add terms 
\begin{equation}    
\lambda \left( \left(  \sum_j (x_{j, k}+ y_{j, k})   \right)\left( \sum_j (x_{j, k}+ y_{j, k})  \right) -\sum_j \left(  x_{j, k}^2+ y_{j, k}^2   \right)  \right)
\end{equation}
to the QUBO Hamiltonian. This is also described as Transformation \#2 in \cite{glover2018tutorial}. On the other hand, the size constraints require implementation with slack variables or use of the unbalanced penalization method \cite{montanez2024unbalanced}. In the numerical experiments, we will mainly consider augmented Lagrangian methods, which can be interpreted as iteratively optimizing the free parameters of the unbalanced penalization formulation.

\subsubsection{Unbalanced penalization description}\label{sec:Unb}
The size inequalities are implemented via the unbalanced penalization method from \cite{montanez2024unbalanced}. In the following, we will write down the corresponding equations. Inequalities \eqref{eq:length} and \eqref{eq:width} are of the form of a general linear system of inequalities $Ax \leq b$. We define the difference between both sides of the inequality as 
\begin{equation} \xi := b-Ax  \end{equation}
with $b$ being a vector with zeros as entries in our case. 
The variable $\xi$ can now be interpreted as excess height $\delta l$ or width $\delta w$.
If one would now minimize $\xi^2$ one would implement an equality instead of the inequality. In our application, this would result in cutting patterns where parts that are cut out will be as close to the ancestor piece as possible.
The idea of unbalanced penalization is now to approximate an energy term $e^{\xi}$, where there is a strong increase for infeasible values, with  quadratic terms like
\begin{equation}
    \xi ^2 -\lambda_{\text{unb.}} \xi,
\end{equation}
where $\lambda_{\text{unb.}}$ needs to be determined via some ablation. While the quadratic term optimizes that the ancestor piece is covered closely with the cut out pieces, the linear term counteracts this. In practice, $\lambda_{\text{unb.}}$ should be chosen at least so large that the placed pieces do not exceed the plate. Multiplying $\xi = \delta w, \delta l $ with the length or width of the ancestor piece yields the area of the cut off pieces at the various stages. Writing everything down for 
\ref{eq:length} yields 
\begin{align}
&(  l_i \sum_t y_{t , i} + \sum _j  l_j x_{i,j} - \sum_t l_t y_{t,i} )^2-\lambda_{\text{unb.}} ( - l_i \sum_t y_{t , i} - \sum _j  l_j x_{i,j} + \sum_t l_t y_{t,i} )  \nonumber \\ 
     & = (  l_i \sum_t y_{t , i} + \sum _j  l_j x_{i,j} - \sum_t l_t y_{t,i} +\frac{\lambda_{\text{unb.}} }{2} )^2-\frac{1}{4}\lambda_{\text{unb.}} ^2. \label{eq:UnbExample} 
\end{align}
Therefore the linear terms effectively lead to an increase of $l_i$ compared to a direct implementation of the inequality with an equality penalty term. If an ancestor is present one can also view this as effectively shrinking the ancestor pieces in the implementation by a fixed length.

\subsubsection{Extension to problem with multiple plates}
If the size of the larger plates is the same everywhere, we can always have the same starting node $s_{1}$. The next starting nodes $s_2,..., s_{P+1}$ each stand for one of the $P$ plates. $y_{s_1, s_i}$ could either be set to identity or one can optimize the number of plates used in that way.
Since the reformulation as a QUBO does not make use of slack variables, we need in total 
\begin{equation}
    \# \textnormal{Var.} = 2( \Tilde{m}+ P +1 )^2 - 2(\Tilde{m} +P+1 )
\end{equation}
binary variables, where $\Tilde{m}= 2m$ if we are free to rotate pieces and $\Tilde{m}= m$ otherwise.
The reason for the subtracted term is that we do not consider $x_{i,i}$ or $y_{i,i}$ variables. One can further reduce the number of variables because of the requirement that the starting pieces are not cut out of required pieces, so that $y_{ s_{1}, s_{i} }=1$, $x_{ s_{1}, s_{i} }=0$  and $x_{ j, s_{i} }= y_{ j, s_{i} }=0$ for $j\neq s_1$. Taking this into account yields the  formula 
\begin{equation}
    \# \textnormal{Var.} = 2(\Tilde{m})^2-2\Tilde{m} +\Tilde{m}+1 = 2(\Tilde{m})^2-\Tilde{m}+1.
\end{equation}
For the case that allows rotation, the number of variables can be even more reduced by making sure that the index for a certain required piece and the same piece rotated do not occur in the same variable.

\subsubsection{Usable Leftovers}
In an industry setting, one can often store the leftover parts for future processing. If this is the case, it is beneficial if the rest pieces are rather large and not cut down too much. Often, they are only stored if they exceed a threshold size. We will argue that this can be better modeled with QUBOs than with 0-1 integer linear programming in this specific parameterization of cutting patterns. First, note that the waste area $A_{\text{rest}}$ could simply be written as a linear term in the objective like 
\begin{equation}
    A_{\text{rest}}= \sum_i \left( w_i (\sum_t l_t y_{t , i}-l_i \sum_t y_{t , i} -  \sum_{j} l_{j} x_{i, j} ) +
l_i (\sum_t w_t x_{t , i}- w_i \sum_t x_{t , i}  -  \sum_{j} w_{j} y_{i, j} )\right).
\end{equation}
However, minimizing or maximizing the total waste area does not take into account how much the rest pieces are cut down. In principle, one might consider 
\begin{equation}
    A_{\text{weighted}}= \sum_i \left( w_i^2 (\sum_t l_t y_{t , i}-l_i \sum_t y_{t , i} -  \sum_{j} l_{j} x_{i, j} )
   +
l_i^2 (\sum_t w_t x_{t , i}- w_i \sum_t x_{t , i}  -  \sum_{j} w_{j} y_{i, j} )\right),\label{eq:Aw}
\end{equation}
so that rest areas that are larger in one dimension influence the result more. The term $A_{\text{weighted}}$ would then be added to the objective with a hyperparameter $\lambda_A$ that is small enough to ensure that no required piece is left out due to the rest piece maximization. In the appendix \ref{sec:CounterEx}, we give an example to showcase that this term does not always prevent unnecessary cutting down of the rest pieces. Intuitively, one could argue that adding only linear terms to the objective is not enough to obtain the best cutting patterns for leftovers as follows: It is not clear if the information that a single entry $x_{i,j}$ equals identity has a positive or negative effect on how usable the leftovers are. For this, one needs to know the values of more entries. One exception may be if $i$ is a starting node. A valid, heuristic strategy for a small amount of cutting stages could be to penalize the stripes in the first cutting stage and hope that this leads to usable leftovers on the right end of the plates. 
We will now further investigate including additional, quadratic terms in the objective.
More specifically, we consider optimization problems like
\begin{equation}
    \argmin_{ A_{\text{size}}\begin{pmatrix}
        x \\ y
    \end{pmatrix} \leq 0, \quad  A_{\text{demand}}\begin{pmatrix}
        x \\ y
    \end{pmatrix} \leq \mathds{1}  }  c_{\text{cost}}^T \begin{pmatrix}
        x \\ y
    \end{pmatrix} - \lambda_{\text{quad}}  A_{\text{quad}}(x,y)\label{eq:usableExact}
\end{equation}
with
\begin{equation}
    A_{\text{quad}}(x,y)= \sum_i \left( w_i^2 (\sum_t l_t y_{t , i}-l_i \sum_t y_{t , i} -  \sum_{j} l_{j} x_{i, j} )^2
   +
l_i^2 (\sum_t w_t x_{t , i}- w_i \sum_t x_{t , i}  -  \sum_{j} w_{j} y_{i, j} )^2\right)
\end{equation}
for usable leftovers.
However, we still want to reformulate this as a QUBO in a way that avoids slack variables with the methods described above. 
This results in
\begin{equation}
    \argmin_{x,y }  c_{\text{cost}}^T \begin{pmatrix}
        x \\ y
    \end{pmatrix} + \lambda_{\text{demand}} \cdot \begin{pmatrix}
        x & y
    \end{pmatrix} Q_{\text{demand}}  \begin{pmatrix}
        x \\ y
    \end{pmatrix} + \lambda_{\text{size}} \cdot\begin{pmatrix}
        x & y
    \end{pmatrix} Q_{\text{size}}(\lambda_{\text{unb.}})  \begin{pmatrix}
        x \\ y
    \end{pmatrix}  - \lambda_{\text{quad}}  A_{\text{quad}}(x,y), \label{eq:usableUnb}
\end{equation}
with $Q_{\text{demand}}$ and $Q_{\text{size}}(\lambda_{\text{unb.}})$ constructed in the way as implied in sections \ref{sec:Unb} and \ref{sec:Ref}. 
More concretely  $ \begin{pmatrix}
        x & y
    \end{pmatrix} Q_{\text{size}}(\lambda_{\text{unb.}})  \begin{pmatrix}
        x \\ y
    \end{pmatrix}$ should amount to a summation of the terms like \eqref{eq:UnbExample} over $i$. 


In the scenario where all required parts have to be cut out, one should implement the demand inequality as equality.

Since the last terms are very similar we will analyze them more closely. Inserting everything in the last two terms yields:
\begin{align}
     \sum_i  &\left( ( \lambda_{\text{size}}-\lambda_{\text{quad}}w_i^2 ) (\sum_t l_t y_{t,i}- l_i \sum_t y_{t , i} - \sum _j  l_j x_{i,j}    )^2 - \lambda_{\text{size}}\lambda_{\text{unb.}} (\sum_t l_t y_{t,i}- l_i \sum_t y_{t , i} - \sum _j  l_j x_{i,j}    ) 
  \right. \nonumber \\ & \left. +
(  \lambda_{\text{size}}-\lambda_{\text{quad}} l_i^2)(\sum_t w_t x_{t , i}- w_i \sum_t x_{t , i}  -  \sum_{j} w_{j} y_{i, j} )^2  - \lambda_{\text{size}}\lambda_{\text{unb.}} (\sum_t w_t x_{t , i}- w_i \sum_t x_{t , i}  -  \sum_{j} w_{j} y_{i, j} ) \right) \\
=\sum_i  &\left( ( \lambda_{\text{size}}-\lambda_{\text{quad}}w_i^2 ) \left(\sum_t l_t y_{t,i}- l_i \sum_t y_{t , i} - \sum _j  l_j x_{i,j}  -
\frac{\lambda_{\text{size}}\lambda_{\text{unb.}}}{2( \lambda_{\text{size}}-\lambda_{\text{quad}}w_i^2 )}
\right)^2
  \right. \nonumber \\ & \left. +
(  \lambda_{\text{size}}-\lambda_{\text{quad}} l_i^2)\left(\sum_t w_t x_{t , i}- w_i \sum_t x_{t , i}  -  \sum_{j} w_{j} y_{i, j} -
\frac{\lambda_{\text{size}}\lambda_{\text{unb.}}}{2( \lambda_{\text{size}}-\lambda_{\text{quad}}l_i^2 )}\right)^2 \right) + \text{const.} \label{eq:Endres}
\end{align}

Since unbalanced penalization is only an approximation, the optimization problem \eqref{eq:usableUnb} will not exactly solve \eqref{eq:usableExact}. However, we get the insight that by making the parameters for the size constraint be dependent on $w_i$ and $l_i$ in the way we see in \eqref{eq:Endres}, we can push the solution in a direction with usable leftovers. 
This comes, however, at the cost of having to tune hyperparameters.
We will provide some numeric evidence with a toy example in Sec.~\ref{sec:ToyExample}.

Furthermore, if the value of the required pieces is simply the area of the rectangles, the first objective term can be merged with the linear unbalanced penalization term, since
\begin{align}
c_{\text{cost}}^T \begin{pmatrix}
        x \\ y
    \end{pmatrix}=  \sum_i \left( l_i (\sum_t w_t x_{t , i}- w_i \sum_t x_{t , i}  -  \sum_{j} w_{j} y_{i, j} ) +
     w_i (\sum_t l_t y_{t,i}- l_i \sum_t y_{t , i} - \sum _j  l_j x_{i,j}    ) \right)-WL.
\end{align}
Including this also in \eqref{eq:Endres} results in

\begin{align}
\sum_i  &\left( ( \lambda_{\text{size}}-\lambda_{\text{quad}}w_i^2 ) \left(\sum_t l_t y_{t,i}- l_i \sum_t y_{t , i} - \sum _j  l_j x_{i,j}  +
\frac{ w_i -\lambda_{ \text{size}}\lambda_{\text{unb.}}}{2( \lambda_{\text{size}}-\lambda_{\text{quad}}w_i^2 )}
\right)^2
  \right. \nonumber \\ & \left. +
(  \lambda_{\text{size}}-\lambda_{\text{quad}} l_i^2)\left(\sum_t w_t x_{t , i}- w_i \sum_t x_{t , i}  -  \sum_{j} w_{j} y_{i, j} +
\frac{l_i-\lambda_{\text{size}}\lambda_{\text{unb.}}}{2( \lambda_{\text{size}}-\lambda_{\text{quad}}l_i^2 )}\right)^2 \right) + \text{const.} \label{eq:EndresWithCost}
\end{align}

An important insight from this formula is that an increase in the unbalanced penalization term is effectively the same as a decrease in the value of the parts. 
Therefore, $\lambda_{\text{unb.}}$ should not be chosen too big. We set
\begin{equation}
    \lambda_{\text{unb.}}\leq \frac{\min \{ l_1,...,l_N,w_1,..., w_N \} }{\lambda_{\text{size}}}
\end{equation}
to make it still in principle beneficial to place required pieces. These bounds could give initial values for the hyperparameters in the unbalanced penalization formulation.

In Sec.~\ref{sec:exp}, we mainly show how these hyperparameters can be tuned iteratively to devise good cutting patterns.
Furthermore, we now look at a toy example to showcase the additional freedom in QUBO formulations.


\subsection{Analytical Toy example}\label{sec:ToyExample}
We want to place $N$ squares of length $a$ in a rectangle of size $(L,W)$. This is done in a way that only $c$ columns are filled with $n(i)>0$ squares for $i=1,..,c$. The formula for the (individually,) squared area of the rest pieces is given by 
\begin{equation}
    W^2( L-ca )^2 + \sum_{i=1}^c a^2(W-n(i)a)^2.\label{eq:Energ}
\end{equation}
With the same coefficient as before, the term that we minimize becomes
\begin{equation}
    (\lambda_{\text{size}} - \lambda_{\text{quad}}W^2)( L-ca  -\frac{\lambda_{\text{size}}\lambda_{\text{unb.}}}{2( \lambda_{\text{size}}-\lambda_{\text{quad}}a^2 )} )^2 + \sum_{i=1}^c (\lambda_{\text{size}} - \lambda_{\text{quad}}a^2)(W-n(i)a-\frac{\lambda_{\text{size}}\lambda_{\text{unb.}}}{2( \lambda_{\text{size}}-\lambda_{\text{quad}}a^2 )})^2.\label{eq:EnergWithParam}
\end{equation}

\begin{figure}
    \centering
\includegraphics[width=0.75\linewidth]{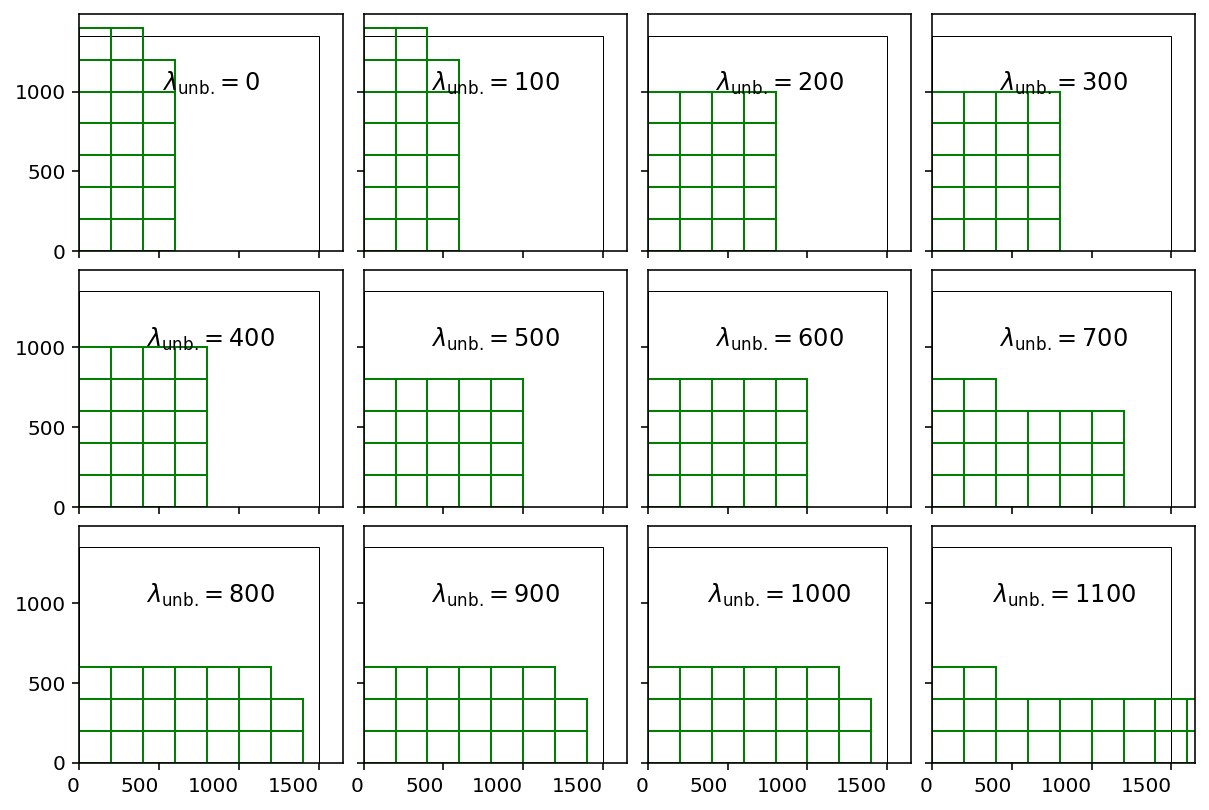}
    \caption{Toy example how $N=30$ squares of size $200$ have to be cut out of a plate to optimize a quadratic energy function depending on an unbalanced penalization parameter $\lambda_{\text{unb.}}$. At $\lambda_{\text{unb.}}=0$, the cutting plan is not feasible, since the pieces slightly exceed the width of the plate. Note that vertical cuts in the first cutting stage are performed for each column of small squares.}
\label{fig:Analyticalillustr}
\end{figure}

Numerical experiments where we always want to place every square reveal that 
with $\lambda_{\text{unb.}}=0$ one should mostly fill the columns even if one exceeds the plate a little. If one has a nonzero $\lambda_{\text{unb.}}$, then $L$ and $W$ in formula \eqref{eq:EnergWithParam} will be effectively decreased. 
Numerical experiments reveal that the minimum is then not one of the extreme cases but somewhere in the middle depending on the value of $\lambda_{\text{unb.}}$. This is illustrated in Fig.~\ref{fig:Analyticalillustr}. While there is likely no single $\lambda_{unb.}$ that yields optimal rest pieces for all settings, we see that the parameter has a large influence in how reasonable the cutting pattern is.

\subsection{Augmented Lagrangian Method}
The unbalanced penalization method has the drawback that one still needs to figure out how to tune the parameters. 
The augmented Lagrangian method as it is described in \cite{sharma2025cutting}, in contrast, works iteratively and entails an update step for the Lagrange parameters. 
The idea is to implement the inequality $Ax-b\leq 0$ by optimizing
\begin{align*}
\mathcal{L}_A(x, \lambda , \mu ) &= f(x) + \sum_{j=1}^{m} \lambda_j \left( b_j - \sum_{i=1}^{n} A_{j,i} x_i \right) 
 + \frac{\mu}{2} \sum_{j=1}^{m} \left( b_j - \sum_{i=1}^{n} A_{j,i} x_i \right)^2
\end{align*}
with iterative updates:
\begin{equation*}
\lambda_j^{(k+1)} = \lambda_j^{(k)} + \mu \left( b_j - \sum_{i=1}^{n} A_{ji} x_i^k \right).
\end{equation*}

The resulting QUBOs can also be obtained from \eqref{eq:usableUnb} with $\lambda_\mathrm{quad}=0$, $\lambda_\mathrm{size}= \mu $ and $\lambda_{\text{unb.}}= \frac{\lambda}{\mu}$. One can also write the energy as
\begin{equation}
    \sum_i \left( \delta l_i - \frac{(\lambda_{\text{unb.}})_i}{2} + \frac{w_i}{2\mu} \right)^2 + \text{Terms with $w,l$ and $x,y$ switched},
\end{equation}
where $\delta l_i$ and $\delta w_i$ denote again residuals in sizes from \eqref{eq:length} or \eqref{eq:width}.
A good choice for a starting point where one wants to have the lengths of the child pieces similar to the lengths of the previous pieces would be 
\begin{equation}
   (\lambda_{\text{unb.}})_i  = 
 \begin{cases}
   \frac{w_i}{2\mu} \quad \text{ for } i\leq N \\
   \frac{l_i}{2\mu} \quad \text{ else }
   \end{cases}.
\end{equation}
The whole algorithm is written up in \ref{alg:cap} with vectors denoted as $x$  referring to all binary decision variables from previous sections, i.e., $\begin{pmatrix} x \\ y \end{pmatrix}$.
The whole algorithm fits well to a stochastic sampler for optimizing the Lagrangian. If the excess heights $\delta l $ and $\delta w$ have values around $0$ for example one can search in the output histogram for feasible solutions with good energy. One can also think about averaging the Lagrangian update over some part of the output histogram. Specifically, if the QUBO solver produces a whole histogram of solution candidates, one can pick the best feasible one as $x_\mathrm{cand}$ in algorithm \ref{alg:cap}.

\begin{algorithm}[H]
\caption{Augmented Lagrangian for 2D cutting stock problems}
\label{alg:cap}
\begin{algorithmic}
\Require{ Problem instance $\{ (l_i, w_i, v_i ) , L , W\}$, \textit{maxiter}}
\Ensure{ Optimized cutting pattern }
\State $\lambda_i \gets 0$
\State $\mu \gets (w_1^2,...,w_m^2, l_1^2,...,l_m^2) $
\State \textit{Set up $A,b$ for inequalities } \eqref{eq:IneqDemand} \textit{and} \eqref{eq:length}
\State $
        x_{\text{best}}  \gets 0 $
    \State  $E_{\text{best}}\gets 0$
\For{$k \gets 1$ to \textit{maxiter}}                    
        \State {$\mathcal{L}_A(x, \lambda , \mu ) \gets c_{\text{cost}}^T 
        x  +\lambda_{\text{demand}} x^TQ_{\text{demand}}x + \sum_{j=1}^{m} \lambda_j \left( b_j - \sum_{i=1}^{n} A_{j,i} x_i \right) 
 + \frac{1}{2} \sum_{j=1}^{m} \mu _j \left( b_j - \sum_{i=1}^{n} A_{j,i} x_i \right)^2$}
     \State {$\lambda_j^{(k+1)} \gets \lambda_j^{(k)} + \mu \cdot \max \{ 0, \left( b_j - \sum_{i=1}^{n} A_{ji} x_i^k \right) \} $}
   \State{$ 
        x_{\text{cand}} 
     \gets \argmin_x \mathcal{L}_A(x, \lambda , \mu )$ (generate samples via QUBO solver) }
\If{Candidate solution $x_{\text{cand}}$ feasible and improve $E_{\text{best}}$}
 \State  $
        x_{\text{best}}  \gets  
        x_{\text{cand} }  $
    \EndIf
    \EndFor 
 \State \Return  Cutting plan corresponding to $
        x_{\text{best}}$
\end{algorithmic}
\end{algorithm}

\section{Numerical Experiments with Simulated Annealing}
\label{sec:exp}

We numerically assess the two formulations introduced above separately: first, we benchmark the 0-1 integer linear programming formulation on larger cutting-stock instances; second, we test the QUBO formulation combined with the augmented Lagrangian method on smaller benchmark instances.

We test the 0-1 integer linear programming formulation on instances from \cite{fayard1998efficient}, which are considered with and without rotation. The achieved values can be viewed in Tables \ref{tab:ResultsCUCW} and \ref{tab:ResultsCUCWTime}. For comparison, the values from the X2D method which is introduced in \cite{velasco2019improved} and an older dynamic programming method DP\_AOG from \cite{morabito2010heuristic} are copied from experimental evaluation in \cite{velasco2019improved}.

To solve the linear programs, we use the HiGHS solver \cite{huangfu2018parallelizing} via highspy with a maximal time of $3$min and a heuristic effort parameter \cite{ergocodeListOptions} of 0.9 if not mentioned otherwise. Although we could not compete with the state of the art methods, here we want to emphasize that we could find optimal solutions on multiple instances. It is also likely that commercial linear programming solver or optimized solver parameters for the HiGHS solver would reduce the times needed for the calculations.

\begin{figure}
    \centering
    \includegraphics[width=0.45\linewidth]{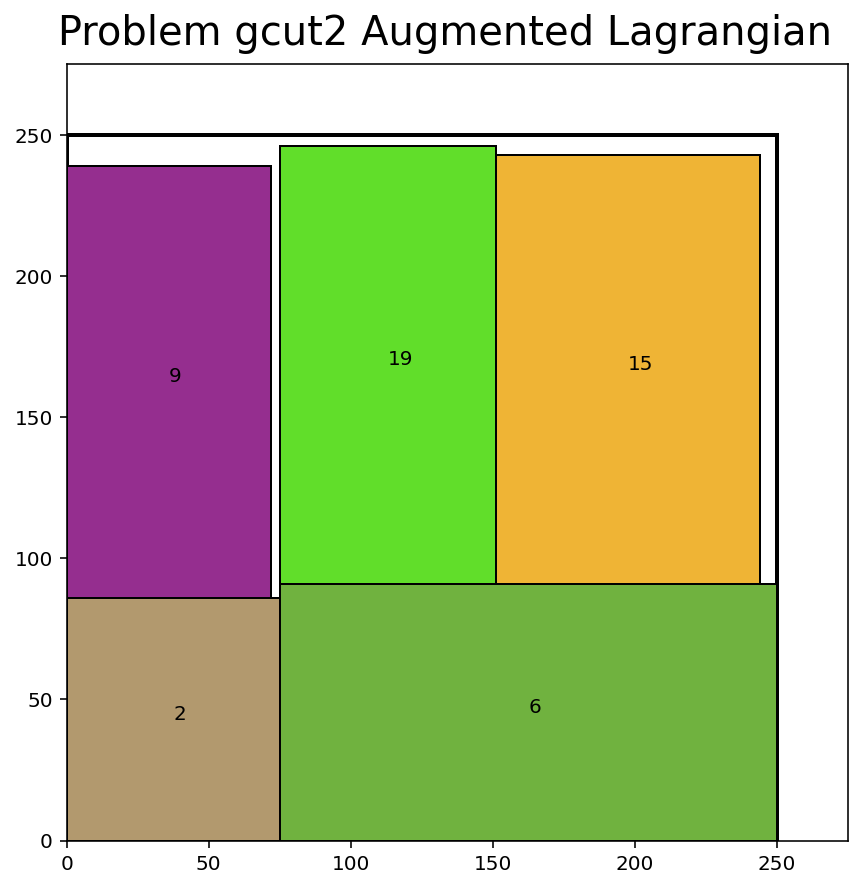}
      \includegraphics[width=0.45\linewidth]{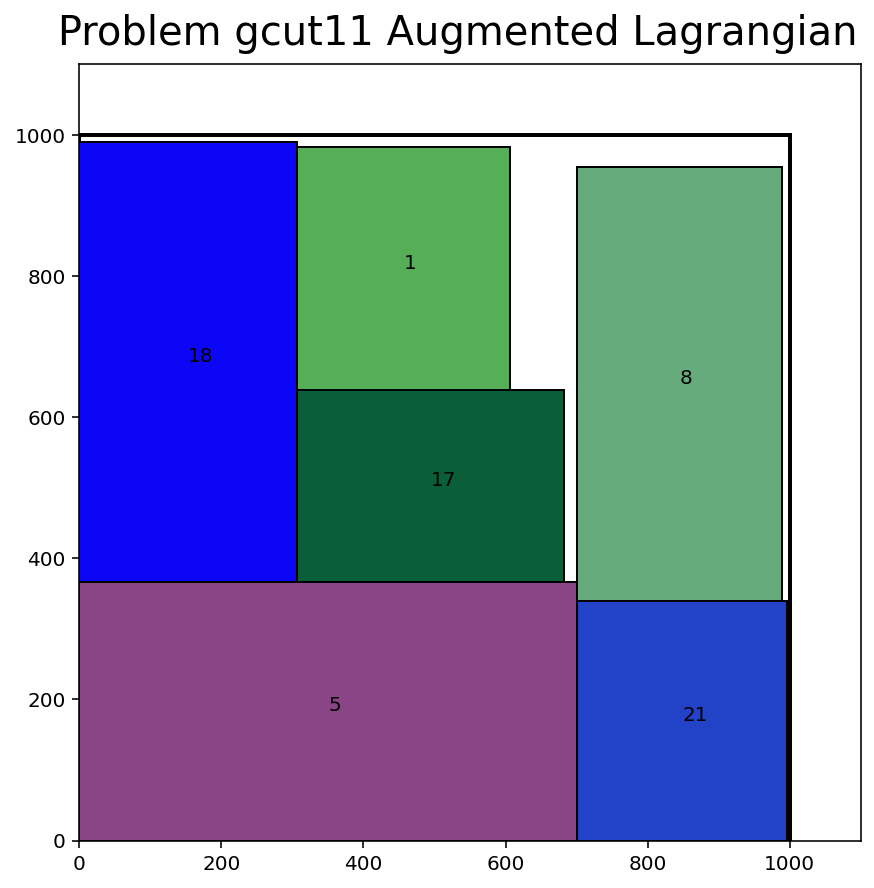}
    \caption{Results from the augmented Lagrangian method 
    for the gcut 2 and 11 instances from \cite{beasley1985algorithms}. The left result yields the optimum of the 0-1 integer linear programming formulation, while the right output was still not optimal. Numbers refer to the required pieces.}
    \label{fig:Teaser}
\end{figure}

\begin{figure}
    \centering
    \includegraphics[width=0.45\linewidth]{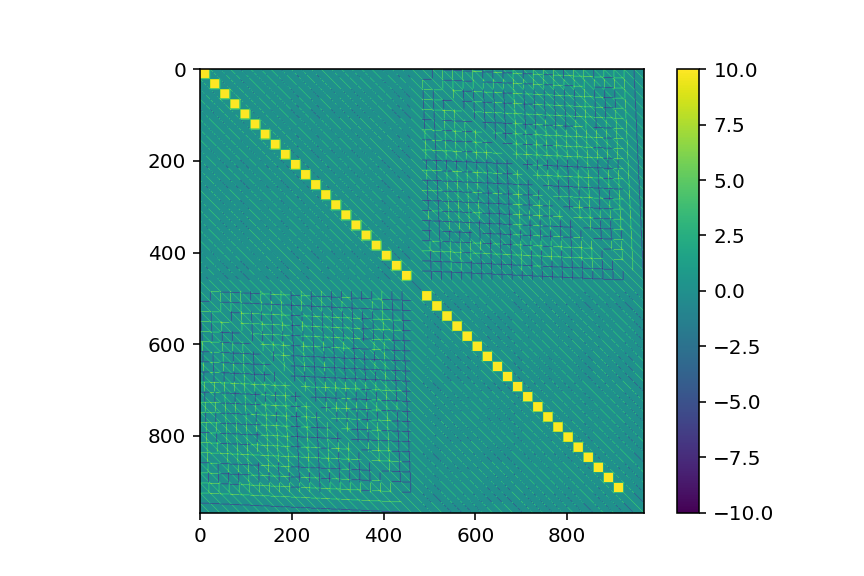}
      \includegraphics[width=0.45\linewidth]{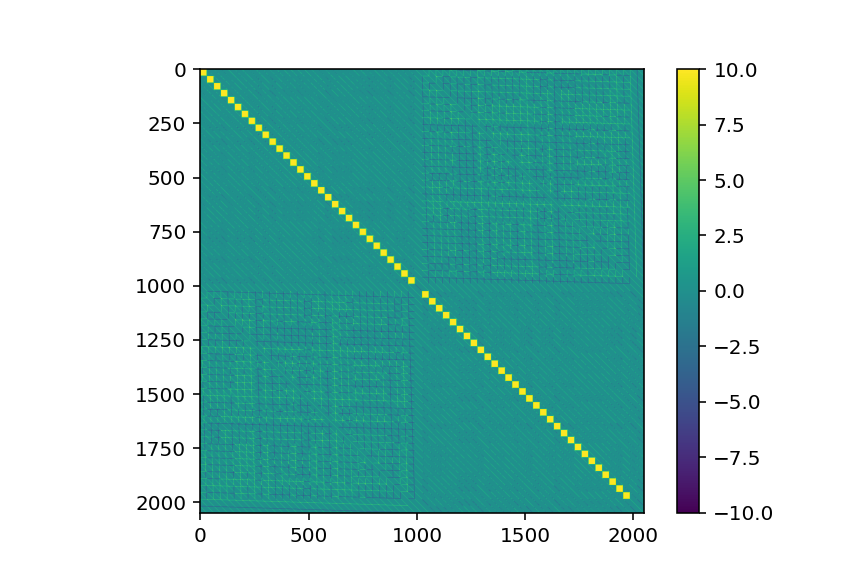}
    \caption{QUBOs from the augmented Lagrangian method  
    for the gcut 2 and 11 instances from \cite{beasley1985algorithms}. Large values in the diagonal are cut off at 10 for better visualization of the off diagonal.}
    \label{fig:QUBOs}
\end{figure}

\begin{figure}
    \centering  \includegraphics[width=0.45\linewidth]{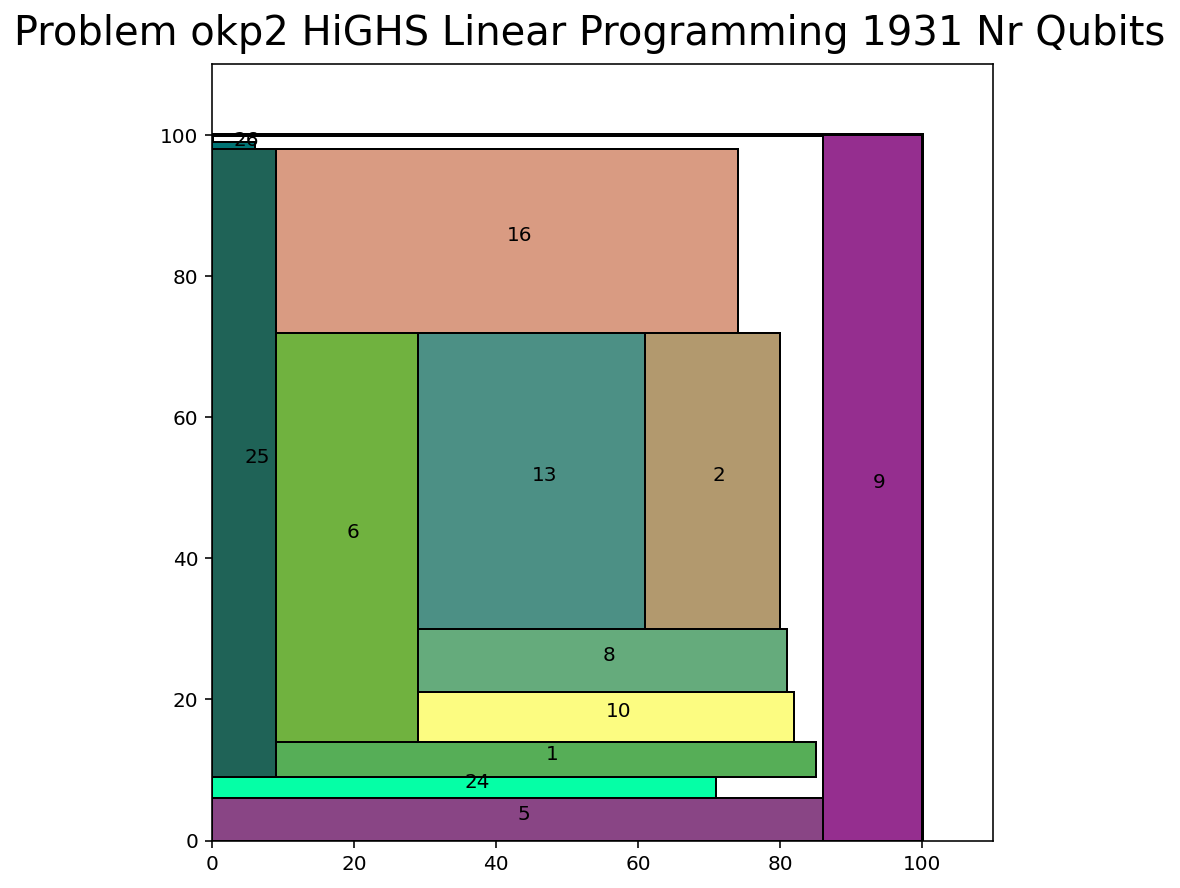}
\includegraphics[width=0.465\linewidth]{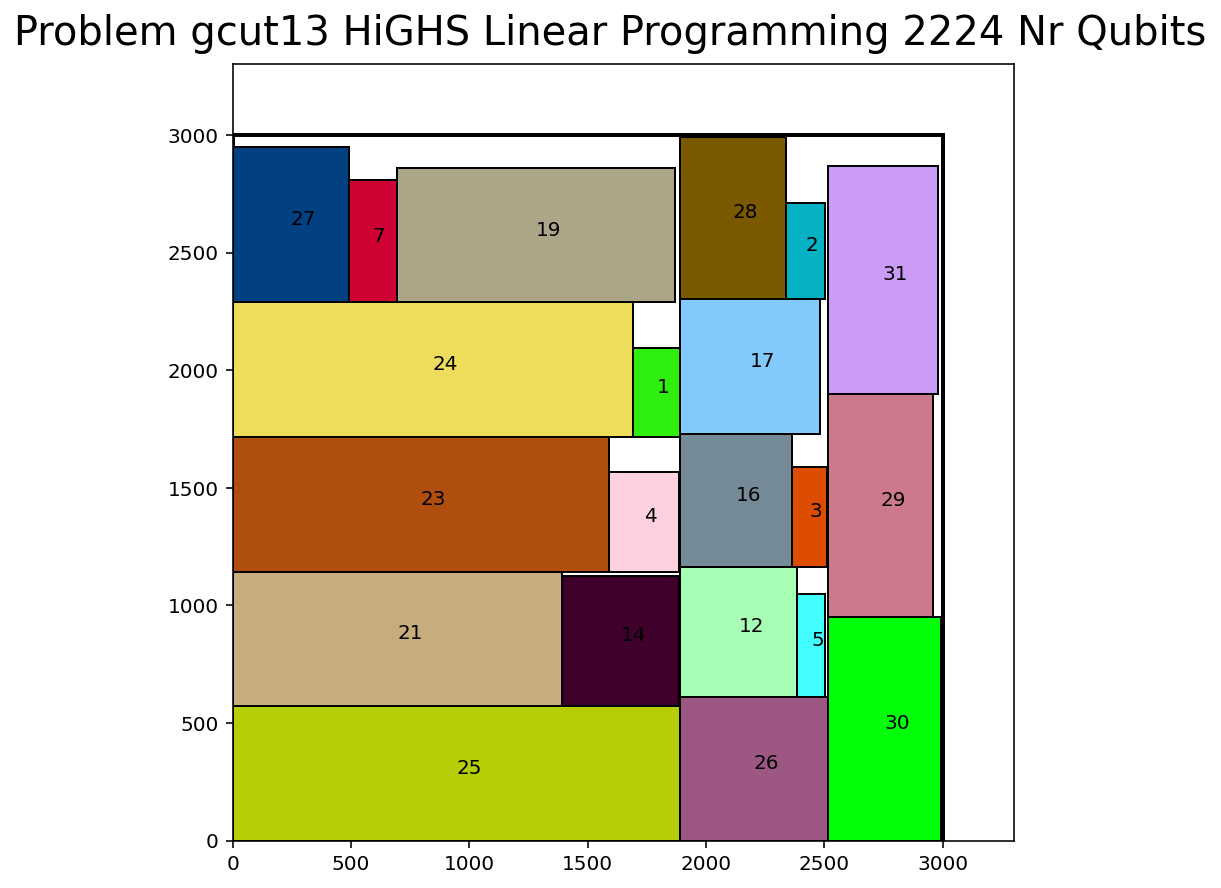}
    \caption{Solution via 0-1 integer linear programming for the instance okp2 from \cite{fekete1997new} and for gcut 13 from \cite{beasley1985algorithms}.}
    \label{fig:negExample}
\end{figure}

We next consider the QUBO formulation
implemented with the augmented Lagrangian method. We investigated the method by executing it on smaller literature instances for $100$ iterations. The individual QUBOs were solved with the neal simulated annealing solver\cite{dwavenealdocsDwavenealx2014}. From the output samples, the best feasible solution over all iterations was determined in a greedy way. Figure~\ref{fig:Teaser}
shows on the left side an exemplary result that was optimally solved with our presented method. 
Figure~\ref{fig:negExample} shows examples for which the QUBO approach did not produce a feasible solution; in these cases, we instead display the solutions obtained with the 0-1 integer linear programming formulation.
In Fig.~\ref{fig:QUBOs}, we additionally show representative QUBO matrices generated during the augmented Lagrangian iterations.

 Table~\ref{tab:SmallGSTable} summarizes the performance of the QUBO approach on the gcut instances \cite{beasley1985algorithms}, reporting whether the optimum of the corresponding linear programming formulation is recovered, together with the QUBO size and the time at which the best solution is first found.
 Instances with the same number of required pieces can result in different numbers of binary variables, since cutting out a bigger piece from a smaller one can be avoided directly.

\begin{table}
    \centering
    \begin{tabular}{|l|l|l|c|c|c|}\hline
 & \multicolumn{2}{|c|}{Optimal}& \multicolumn{3}{|c|}{Restricted Problem LP}\\
    \hline
Instance & Rot. &No Rot. &     Rot. & No Rot. &  Rot.less Heuristic\\   
\hline
CU1: & 12,500&12,330& 12,027& 12,312& 12,336\\ \hline
CU2: & 26,200&26,100& 24,408& \textbf{26,100}& 25,709\\ \hline
CU10: & 779,239&773,772& 694,388& 748,546& 735,891\\ \hline
CW1: & 6766&6402& \textbf{ 6766}& \textbf{6402}&  \textbf{ 6766}  \\ \hline
CW2: & 5689&5354& 5540& \textbf{5354}& 5540\\ \hline
CW3: & 5744&5689& \textbf{5744}& 5674& \textbf{5744}\\ \hline
CW4: & 7496&6175& \textbf{7496}& 6158& \textbf{7496}\\  \hline
CW5: & 11,659&11,659&  \textbf{11,659}& 11,644&  \textbf{ 11,659}\\ \hline
CW6: & 13,203&12,923& 13,027& 12,635& 13,027\\ \hline
CW7: & 10,880&9898& 10,204& 9484& 10,204\\ \hline
CW8: & 4736&4605& 4464& 4504& 4580\\ \hline
CW9: & 11,479&10,748& 1462& \textbf{10,748}& 11,224\\ \hline
CW10: & 6835&6515& 6301& 6016& 6306\\ \hline
CW11: & 6784&6321& 6188& 5822& 6188 \\ \hline
    \end{tabular}

    \caption{
    Objective values obtained with the linprog formulation for the CU and CW benchmark instances, with and without piece rotation. Literature values for the unrestricted problem are reported for comparison in the 'Optimal' column. 
    In the last column, the mip heuristic effort parameter was set to $0.1$ and the maximal time was set to $30$ min. The optimal values are from the non-restricted problem. Bold indicates optimal energy for our experiments.}
    \label{tab:ResultsCUCW}
\end{table}

\begin{table}
    \centering
    \begin{tabular}{|l|l|l|l|c|c|c|}\hline
 & \multicolumn{3}{|c|}{Optimal}& \multicolumn{3}{|c|}{Restricted Problem LP}\\
    \hline
Instance &Rot. X2D in s &No Rot. X2D in s&No Rot.  DP\_AOG in s&     Rot. in s& No Rot. in s&  Rot.less Heuristic in min\\   
\hline
CU1: &0.01&0.01&0.5& 180& 180& 30\\ \hline
CU2: &0.02&0.01&1.4& 180& \textbf{180}& 30\\ \hline
CU10: &0.06&14.36&1800& 181& 180& 29.6\\ \hline
CW1: &0.21&0.01&25.6& \textbf{ 180}& \textbf{6.5}&  \textbf{ 19.0}\\ \hline
CW2: &0.18&0.62&1118.8& 180& \textbf{23.6}& 5.2\\ \hline
CW3: &4.83&0.03&26.9& \textbf{181}& 4.0& \textbf{10.7}\\ \hline
CW4: &0.15&0.13&46.8& \textbf{180}& 180& \textbf{6.68}\\  \hline
CW5: &0.12&0.02&1268.5&  \textbf{180}& 31.7&  \textbf{ 6.4}\\ \hline
CW6: &5.64&0.05&890.6& 182& 61.3& 30\\ \hline
CW7: &0.01&0.01&10.7& 182& 147.9& 30\\ \hline
CW8: &0.10&0.06&340.5& 182& 180& 30\\ \hline
CW9: &0.12&0.04&121.3& 180& \textbf{27.2}& 30\\ \hline
CW10: &0.24&0.04&182.2& 181& 33.5& 29.5\\ \hline
CW11: &0.41&0.19&1375& 180& 69.5& 18.1\\ \hline
    \end{tabular}
    \caption{Computational time results needed for the linprog formulation on the CU and CW instances. In the last column, the MIP heuristic effort parameter was set to $0.1$ and the maximal time was set to $30$ min. Otherwise, we had a MIP heuristic effort parameter of $0.9$ and a time restriction of $180$s. Achieving optimal energy values is indicated by bold numbers.}
    \label{tab:ResultsCUCWTime}
\end{table}

\begin{table}
    \centering
    \begin{tabular}{|c|c|c|c|c|}
     \hline
        Instance name & Optimality & $\Tilde{m}$ & QUBO size  & Time until best solution occured in min\\  \hline
        gcut1 & yes & 10 & 245 & 0.01 \\ \hline
         gcut2 & yes &  20&  888 & 1.43 \\ \hline
         gcut3 &  no & 30 & 1931 & 6.54 \\ \hline
         gcut4 & no &  50 & 5219 & 217.07 \\ \hline
         gcut5 & yes & 10 & 244 & 0.35 \\ \hline
         gcut6 & yes & 20 & 886 & 2.17 \\ \hline
         gcut7 & yes  & 30  & 1929 & 10.27\\ \hline
         gcut8 & no & 50 & 5214 & 200.17\\ \hline
        gcut9 & yes  & 10  & 244   & 0.02\\ \hline
        gcut10 & no & 20 & 884 & 0.58 \\ \hline
        gcut11 & no & 30 & 1925 & 4.00 \\ \hline
         gcut12  & no &  50  & 5208 & 405.5\\ \hline
        gcut13  &  no& 32 &  2224 & 0.02 \\ \hline
    \end{tabular}
    \caption{List of benchmark instances for which the augmented Lagrangian method did or did not find the optimal solution.}
    \label{tab:SmallGSTable}
\end{table}

\section{Scaling}
\label{sec:scaling}
As we have seen, the above formulation requires no slack variables and the number of qubits scales quadratically with the number of required pieces. Another quantity of interest is the amount of off-diagonal elements in the QUBO matrix. There are only couplings between two binary variables if they refer to the same piece. Let $x_{i,j}$ and $x_{k,l}$ denote two binary entries with the corresponding indices. We have some interaction if $k=i$ or $l=j$. If one is switched to a $y-$variable, we also have interaction for $i=l$ or $j=k$.
If one considers the Characteristic of the starting indices, the following formula for the number of edges can be obtained
\begin{align}
    &2(m^2-m)\frac{4(m-2)}{2} +m(\frac{m-1}{2}+2(m-1))+m \nonumber\\
    &= 4 m^3 - \frac{19}{2} m^2 +\frac{13}{2} m \,.
\end{align}

The fraction of entries where we have interaction $I$ is therefore
\begin{equation}
    I= \frac{ 4 m^3 - \frac{19}{2} m^2 +\frac{13}{2} m }{(2(m^2-m)+m+1)^2}, 
\end{equation}
which goes to zero in the limit $m\to \infty$ with asymptotic behavior $ \sim \frac{1}{m}$.
Due to the size inequalities, a single binary variables has nonzero coupling values to $2(N-2)$ other variables.
According to \cite{mehta2022hardness}, the Hamming distance between local optima is an important quantity for estimating how hard a problem is for quantum annealing. If one wants to only remove leave indices, the Hamming distance only changes by one. However, to change out a piece that is generated in an intermediate cutting stage in some branch often at least two entries need to be changed. Intuitively, large Hamming distances can occur if all parts of a branch need to be changed. 

\begin{figure}
    \centering
    \includegraphics[width=0.5\linewidth]{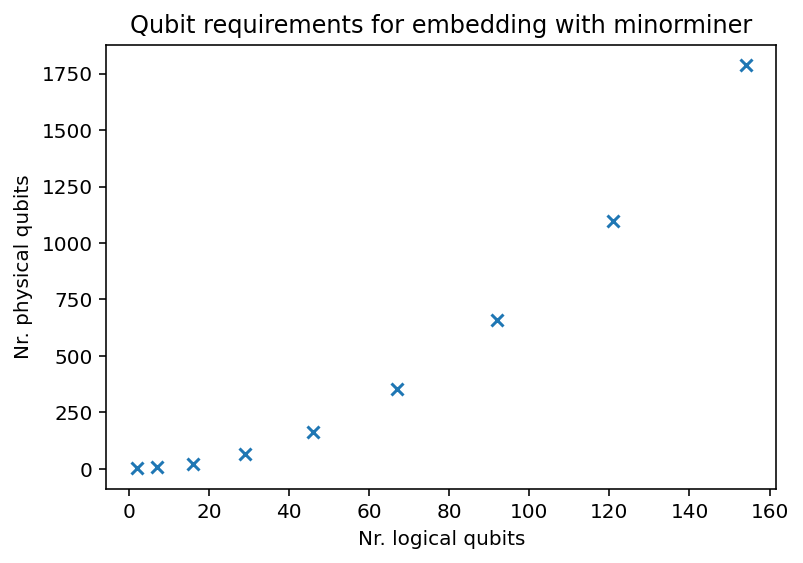}
    \caption{Relationship between the number of logical qubits and the number of physical qubits after embedding on D-Waves Pegasus graph via the minorminers \cite{cai2014practical,minorminerdocsMinorminerx2014} default method.}
    \label{fig:dwaveEmb}
\end{figure}

Another aspect about the scaling behavior worth researching is given by the qubit connectivity requirements. 
In superconducting devices, hardware constraints that lead to limited connectivity have been identified early on as a major bottleneck~\cite{Katzgraber}, and various workarounds such as the parity encoding~\cite{Lechner2015}, minor embeddings~\cite{cai2014practical} or specialized, iterative methods~\cite{benkner2025compensating} have been developed. The desire to increase the connectivity has led to the design of chips such as the D-Wave quantum annealing Pegasus hardware \cite{boothby2020next}. 
In Fig.~\ref{fig:dwaveEmb}, the relation between the numbers of logical and physical qubits after an embedding to the Pegasus graph is depicted. One can see a strong increase in the number of physical qubits required as the problem size increases. 

In contrast, as discussed in Sec.~\ref{sec:PhysIntro} the ions in a single trap have all-to-all connectivity. However, scaling to large problem instances requires the use of multiple traps or distributed trapping sites. Thus, a suitable comparison of the number of required physical qubits with the situation on ion traps is given by the concept of distributed quantum computing. A common scenario is that one has multiple registers with all-to-all connectivity, between which the ions are moved. This shuttling between registers may result in principle in arbitrary connectivity but comes with drawbacks like a larger shuttling time or errors because of excitations \cite{kaushal2020shuttling}. It is an active research area to devise strategies how to handle couplings across registers \cite{rajabi2026distributed}.

\begin{figure*}
\centering
    \begin{subfigure}{0.45\linewidth}
        \includegraphics[width=1.2\linewidth]{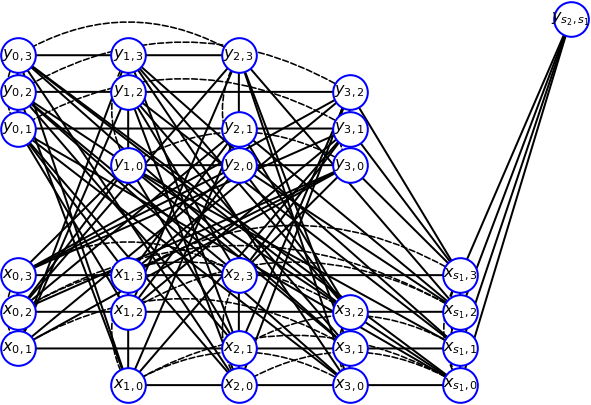}
        \label{fig:2a}
    \end{subfigure}
    \begin{subfigure}{0.45\linewidth}
    \includegraphics[width=0.73\linewidth]{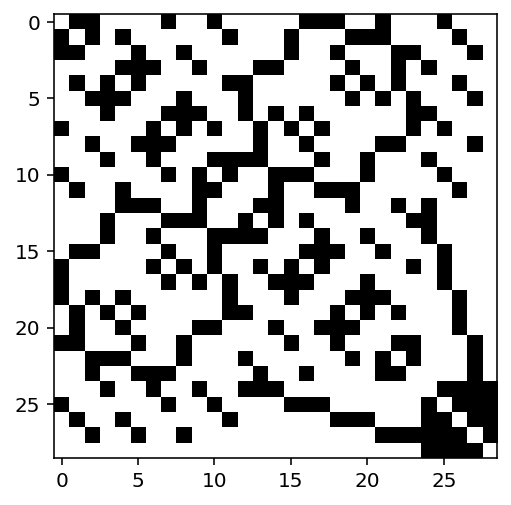}
        \label{fig:2b}
    \end{subfigure}
\caption{Connectivity graph of the QUBOs and illustration corresponding to the adjacency matrix for $m=4$.}
\end{figure*}

\section{Simulation of Quantum Annealing protocols}
\label{sec:Simulating}

To investigate the application of quantum annealing to the QUBO formulation introduced above, we simulate the annealing dynamics for three representative cutting-stock instances involving two or three required pieces. 
The instances are chosen such that the corresponding QUBOs remain sufficiently small for an exact simulation of the quantum dynamics, resulting in two four-qubit problems and one nine-qubit problem. The two four-qubit instances involve two required pieces: in one case both pieces can be placed, while in the other only one can be accommodated. The nine-qubit instance involves three required pieces, all of which can be placed. For the four-qubit instance in which only one piece can be placed, we consider the QUBOs obtained at different iterations $k$ of the augmented Lagrangian method. The QUBO formulation changes with each $k$ value for $k\leq3$. Only for $k\geq3$ does the ground state of the QUBO problem encode the solution to the cutting stock problem instance. For the other four-qubit instance and the nine-qubit instance, the QUBO remains unchanged from $k=0$ onward, so that all subsequent iterations correspond to the same QUBO. The dimensions and QUBO coefficients of these instances are provided in the repository~\cite{github}.
For each instance, the QUBO cost function is mapped onto an Ising Hamiltonian $H_\mathrm{cost}$, while a transverse-field Hamiltonian $H_\mathrm{drive}$ is used as the initial driver Hamiltonian. We consider a linear annealing protocol:

\begin{equation}
    H(s)=(1-s)H_\mathrm{drive}+sH_\mathrm{cost}, \qquad s=t/T,
\end{equation}
where $t$ is the evolution time and $T$ is the total annealing time. 
The system is initialized in the ground state $|\psi(0)\rangle$ of $H_\mathrm{drive}$ and evolved according to the time-dependent Schrödinger equation. Since the considered instances are small, the evolution is computed using a Magnus expansion of the time evolution operator (see also Ref.~\cite{nagies2026practical}). At the end of the protocol, we evaluate the fidelity with respect to the ground state $|\psi_{\mathrm{GS}}\rangle$ of the final problem Hamiltonian,
\begin{equation}
    \mathcal{F}(T)=|\langle \psi_{\mathrm{GS}}|\psi(T)\rangle|^2.
\end{equation}
The fidelity therefore gives the probability of obtaining the optimal solution at the end of the annealing protocol.
\begin{figure}
    \centering
    \includegraphics[width=\linewidth]{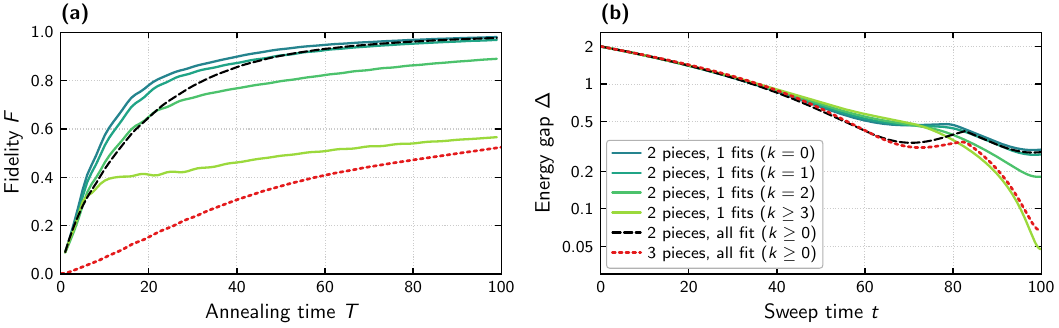}
    \caption{Quantum annealing dynamics for the three representative cutting-stock instances.(a) Final ground-state fidelity $\mathcal{F}$ as a function of the total annealing time $T$  expressed in units of the inverse maximum coupling $J_{\max}^{-1}$. For the four-qubit instance in which only one of the two required pieces can be placed, different curves correspond to different iterations $k$ of the augmented Lagrangian method. For the other four-qubit instance and the nine-qubit instance, the QUBO remains unchanged from $k=0$ onward. (b) Instantaneous energy gap $\Delta=E_1-E_0$ along the corresponding annealing protocols as a function of the normalized annealing parameter $s=t/T$.}
    \label{fig:Sim}
\end{figure}

Figure~\ref{fig:Sim}(a) shows the final fidelity as a function of the total annealing time $T$ in units of the inverse maximal two-body coupling strength for all the considered QUBO instances. The fidelity generally increases with $T$, approaching unity for sufficiently slow protocols in most of the four-qubit cases, although the convergence rate strongly depends on the specific QUBO instance. In contrast, the four-qubit instance,  where only one piece fits, at $k\geq3$ exhibits a substantially slower convergence, comparable to that observed for the nine-qubit problem. For both cases, the final fidelity remains around $0.5$ even for the longest annealing time considered, indicating a more challenging ground-state preparation. This already indicates that the difficulty of the annealing dynamics cannot be inferred from the number of qubits alone, but depends strongly on the specific QUBO generated by the optimization procedure.

For the four-qubit cutting stock instance whose QUBO formulation changes across the augmented Lagrangian iterations, increasing k makes the corresponding quantum annealing problem progressively more difficult. However, a $k$ value of at least 3 is required to encode the correct solution to the original problem in the ground state.

To further illustrate the relation between problem instance and performance of the annealing protocol, Fig.~\ref{fig:Sim}(b) shows the instantaneous energy gap
\begin{equation}
    \Delta(s)=E_1(s)-E_0(s),
\end{equation}
where $E_0(s)$ and $E_1(s)$ are the instantaneous ground-state and first-excited-state energies of $H(s)$, respectively. The smallest energy gap occurs towards the end of the annealing protocol for all considered problems; additionally we observe a local minimum in the energy gap for some of the problem instances. In particular, for the four-qubit instance analyzed at different augmented-Lagrangian iterations, the minimum gap decreases as $k$ increases. This behavior is consistent with the slower convergence of the fidelity observed in Fig.~\ref{fig:Sim}(a).

\section{Conclusions}
\label{sec:conclusion}
The 2D cutting stock problem with Guillotine cuts is a challenging problem of significant industrial relevance. In this work, we introduced a 0-1 integer linear programming formulation to tackle a restricted case. We also provided a reformulation as a QUBO problem. While it has the drawback that the various inequality constraints need to be handled using, e.g., the augmented Lagrangian method, we speculate the QUBO formulation to have potential benefits in addressing the generation of usable leftovers and in generalization to the unrestricted case. Our numerical experiments can be seen as a proof of concept for our formulation. While the problems that we can tackle with the 0-1 integer linear programming formulation are still larger than those that can be tackled with QUBO problems, we could optimally solve smaller literature instances via QUBOs and achieve reasonably good cutting patterns on many more problems.
We further investigated the quantum annealing dynamics of representative cutting stock QUBO instances. The results show that the annealing difficulty depends not only on the problem size but also on the specific QUBO formulation.

A promising goal for future research is to improve the 0-1 integer linear programming formulation and try to make it competitive. One could also investigate in more detail how to realize the QUBO optimization on a quantum annealing device. Although the goal of developing QUBO models to address the 2D cutting stock problem was achieved, this work is inconclusive regarding the question of whether these problems will be a suitable large-scale use case for quantum annealing. In particular, it seems to be difficult to phrase the problem in the general form as a single QUBO optimization. In classical computing, there are various ways how recursion and dynamic programming can help to address 2D cutting stock problems. It seems an interesting research question whether one can integrate these into a hybrid quantum--classical protocol. 
 
We hope that this work inspires other researchers to think about how QUBOs can be applied to solve complicated cutting stock problems, and in this way bring new ideas to the operation research community, gain new insights into hybrid decomposition techniques, or develop interesting benchmark instances for quantum annealing hardware.


\section*{Code Availability}
Code and data used in this work are available: \url{https://github.com/MSeelbach/QUBOs_for_2DCS}

\section*{Acknowledgments}
We thank Marc Goerigk for useful discussions. 
Furthermore, Jonas B. Weber, Tim Fabian Korzeniowski and Jan Leisse provided valuable insights in industrial use cases of 2D cutting stock and 3D nesting problems. 
The work reported in this publication is based on a project that was funded by the German Federal Ministry for Education and Research under the funding reference number 13N16437. 

\appendix

\section{Error from the restriction}
As already discussed, we restrict our cutting patterns to those ones where we cut stripes of widths of required pieces in each step and consider the remaining parts as rest. A more general case would be able to exclude an intermediate piece from a cutting stage but cut it down further in later cutting stages.

\begin{figure}
    \includegraphics[width=\textwidth]{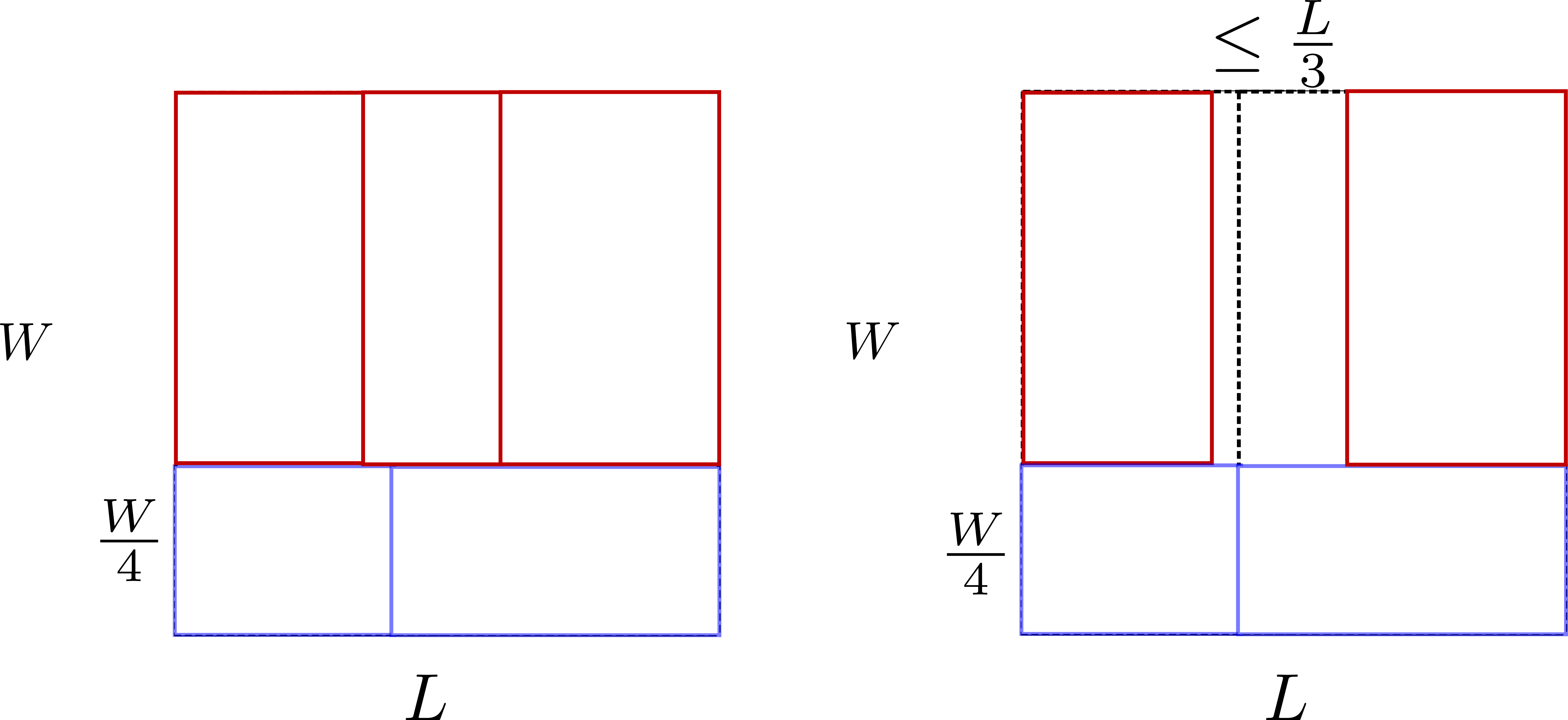}
    \caption{Illustration of a suboptimal solution due to discarding pieces immediately if they are not cut in the previous cutting stage. We enforce that the first cut is vertical in the right drawing for demonstration purpose. More general models can make no cut at all in one stage or make a cut with length $L$ to achieve the result on the left side.}
    \label{fig:restr}
\end{figure}

To lower bound the potential loss in material area, we present an example with 5 required pieces that are cut out without considering rotations. Since we are able to scale the problem, we consider length and width of the plate as 1. The pieces we consider can be obtained by first performing a horizontal cut to get one stripe with height $e$ and one with height $1-e$. After this, there will be some vertical cuts to get to five pieces. If one would start with horizontal cuts, one could cut out everything without loss.
However, since we are interested in the restrictions of our model, we consider the case where we have to perform vertical cuts first. If one piece has the length of the plate, one can pretend to cut out a stripe vertically that is just the whole plate. In this case, the restriction does not prevent the optimal cutting pattern. In contrast, if two pieces are cut out from the first stripe and three from the second, the solution will be suboptimal. The smallest piece on the stripe with three pieces has to be discarded or, if the biggest piece is present at the stripe with three pieces, the smaller piece at the stripe with two pieces will be left out. A failure case like this is illustrated in Fig.~\ref{fig:restr}. In such a scenario, only $\frac{3}{4}$ of the area can be used properly. 

\section{Promising linear term for usable leftovers not working}
\label{sec:CounterEx}
\begin{figure}[H]
    \centering    \includegraphics[width=\linewidth]{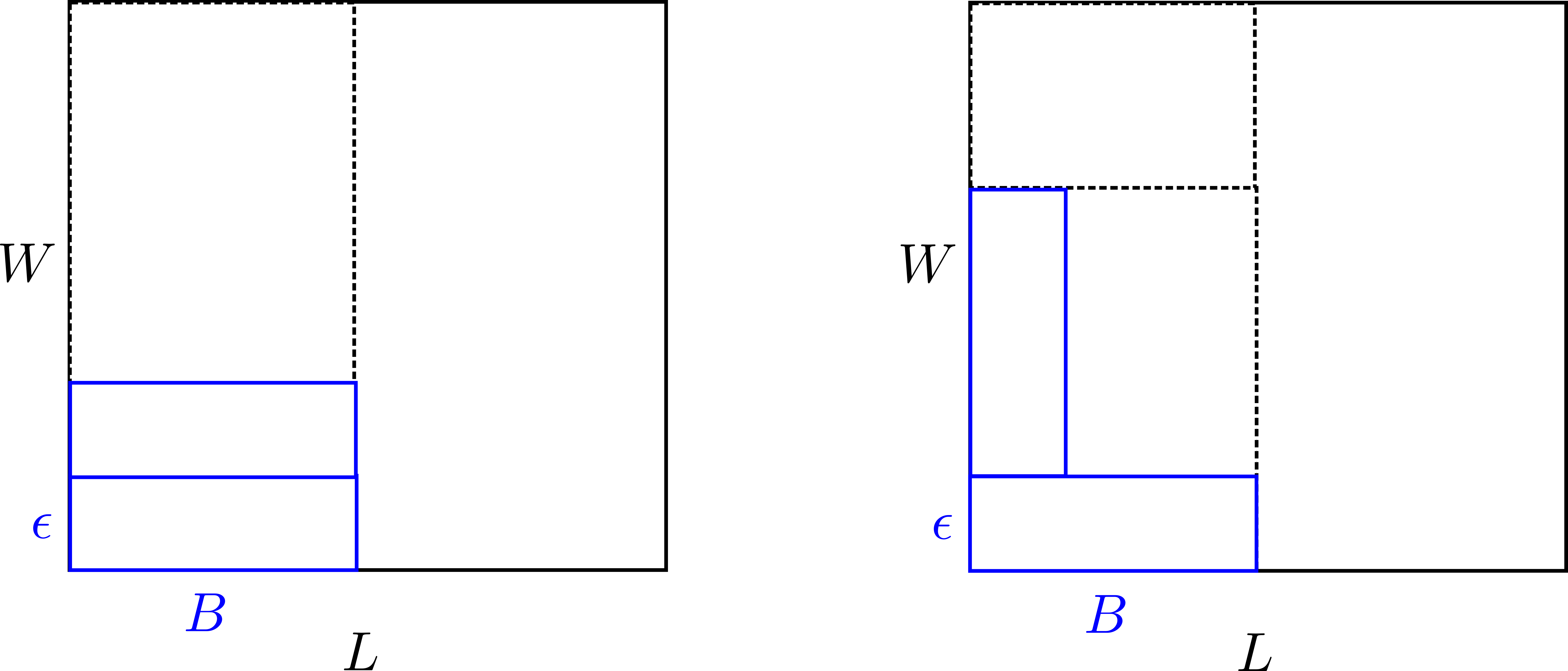}
    \caption{Illustration that the proposed linear term does not prevent all unnecessary cuts.}
    \label{fig:UsableLeftover}
\end{figure}

Consider the cutting patterns illustrated in Fig.~\ref{fig:UsableLeftover}. We now show that $A_{\text{weighted}}$ from \eqref{eq:Aw} is the same for both, although in the below cutting pattern there is an unnecessary cut. Writing down the terms yields
\begin{align}
    &W^2(L-B)+ B^2(W-2\epsilon) \nonumber \\
    =&   W^2(L-B)+ B^2(W-\epsilon-B) + B^2(B- \epsilon).
\end{align}
The quadratic term on the other hand shows that the cut is unnecessary:
\begin{align}
    &W^2(L-B)^2+ B^2(W-2\epsilon)^2 \nonumber \\
    \geq &   W^2(L-B)^2+ B^2(W-\epsilon-B)^2 + B^2(B- \epsilon)^2.
\end{align}

\subsection{Linear inequality constraints for the unrestricted case}
To extend the formulation to the unrestricted problem, we introduce $N^2$ additional variables $z_{t,s}$. If the variable $z_{t,s}$ equals identity, then the pieces $t$ and $s$ are joined in the sense that they determine the bound in a size constraint together. Whether they are stacked vertically or horizontally is determined by the other variables. The demand inequality in \eqref{eq:IneqDemand} can be split into several inequalities and modified to 
\begin{align}
    \forall s,t,k  \quad x_{s,k} +x_{t,k}+y_{s,k} +y_{t,k}  & \leq 1+ z_{s,t}.
\end{align}
Furthermore, if pieces are joined together they should have previously been cut out in the same cutting stage: 
\begin{align}
    \forall i,t,s  \quad  z_{s,t} & \leq 1+ x_{i,t} -x_{i,s}  \\
     z_{s,t} & \leq 1+ x_{i,s} -x_{i,t}  \\
      z_{s,t} & \leq 1+ y_{i,t} -y_{i,s}  \\
     z_{s,t} & \leq 1+ y_{i,s} -y_{i,t}.  
\end{align}
Also, pieces that will be cut out in the next cutting stage should be the same for both of the joined together pieces: 
\begin{align}
    \forall i,t,s  \quad  z_{s,t} & \leq 1+ x_{t,i} -x_{s,i}  \\
     z_{s,t} & \leq 1+ x_{s,i} -x_{t,i}  \\
      z_{s,t} & \leq 1+ y_{t,i} -y_{s,i}  \\
     z_{s,t} & \leq 1+ y_{s,i} -y_{t,i}.  
\end{align}

Since typically either $x_{k,i}$ or $y_{k,i}$ is zero we can combine some of the inequalities:
\begin{align}
    \forall i,t,s  \quad - x_{i,t} +x_{i,s} - y_{i,t} +y_{i,s} +z_{s,t} & \leq 1   \\
  - x_{i,s} +x_{i,t}-y_{i,s} +y_{i,t} +  z_{s,t} & \leq 1   
\end{align}
and
\begin{align}
    \forall i,t,s  \quad - x_{t,i} +x_{s,i}- y_{t,i} +y_{s,i}+ z_{s,t} & \leq 1  \\
   - x_{s,i} +x_{t,i}- y_{s,i} +y_{t,i} + z_{s,t} & \leq 1  \,.
\end{align}
Since multiple pieces can now refer to the same next piece, e.g., $\sum_j x_{j, k}+ y_{j, k} $ can be greater than one, the linear objective also has to change. The idea is to use $z_{k,k}$ as a variable that indicates if the $k$-th piece is placed somewhere. For this, we include the inequalities
\begin{equation}
  \forall k \in \{ 1,...,N \} \quad  z_{k,k} \leq \sum_j x_{j, k}+ y_{j, k}
\end{equation}
and change the objective to
\begin{equation}
    -\sum_k z_{k,k} v_{k}\,.
\end{equation}

Furthermore, an idea to help the ILP-solver get good solutions was to add some restriction as
\begin{equation}
    \sum_{i< j} z_{i,j}\leq 5.
\end{equation}
Nevertheless, numerical evaluations were not that promising.
While this approach can solve a more general cutting stock problem case, it has the drawback that it requires an excessive amount of inequality constraints. Therefore, it was not possible to solve the previous investigated literature instances from Table \ref{tab:ResultsCUCW} optimally within an hour.

\subsection{Quadratic inequality constraints for the unrestricted case}

In this section we give a short description how one could extend our model to the unrestricted case at the cost of additional variables and quadratic terms in the inequalities.
We introduce variables $z_{k,l}$ into the formulation. They indicate that part $k$ and  $l$ will occur together in the sense that a vertical or horizontal cut with the sum of their dimensions will be performed. We use the convention that $z_{k,l}$ can only be one if $l$ is larger $k$. From $l$, there should not be further connections in this case: 
\begin{equation}
 \forall i,k \quad   x_{l,i}+ y_{l,i}   \leq 1-z_{k,l}
\end{equation}

Inequality \ref{eq:IneqDemand} now changes to
\begin{equation}
\forall k \in \{1,...,N\} \quad \sum_j x_{j, k}+ y_{j, k} + z_ {j,k}  \leq 1. 
\end{equation}

The new size constraints are obtained by substituting
$ l_t y_{t,i}\rightarrow l_t y_{t,i} +y_{t,i} ( \sum_ k l_k z_{t,k} )  $
and 
$ l_i x_{i,j}\rightarrow l_j x_{i,j} +x_{i,j} ( \sum_ k l_k z_{j,k} )  $
in the inequality \ref{eq:length} and analogous terms in \ref{eq:width}. This results in 
\begin{equation}
 l_i \sum_t y_{t , i} +  \sum_{j}\left( l_j x_{i,j} +x_{i,j} ( \sum_ k l_k z_{j,k} ) \right) \leq 
 \sum_t \left( l_t y_{t,i} +y_{t,i} ( \sum_ k l_k z_{t,k} ) \right) \label{eq:lengthUnr}
\end{equation}
and 
\begin{equation}
 w_i \sum_t x_{t , i}  +  \sum_{j} \left( w_j y_{i,j} +y_{i,j} ( \sum_ k w_k z_{j,k} )\right)\leq 
 \sum_t\left( w_t x_{t,i} +x_{t,i} ( \sum_ k w_k z_{t,k} ) \right)\,. \label{eq:widthUnr}
\end{equation}

For the objective, we use the term
\begin{equation}
    \sum_ { k, i } (x_{k,i}+ y_{k,i} +z_{k,i}  )  v_{i}.
\end{equation}
If we would now use unbalanced penalization to realize the inequalities, we would get terms of degree four. Advantages of higher order terms have been discussed in the literature \cite{nagies2025boosting,stein2023evidence}, but in practice they still remain challenging to realize on hardware \cite{nagies2024role}. We consider finding other QUBO formulation that address the most general 2D cutting stock problem with Guillotine cuts an open problem. In principle, it would also be possible to introduce a binary variable for each rectangular piece that could be in some way generated throughout the cutting process. However, this will lead to unreasonably large optimization problems in particular if the number of cutting stages is not restricted.

\bibliography{apssamp}

\end{document}